\documentclass[aps,pre,twocolumn,showpacs]{revtex4}
\usepackage{graphicx,epsfig}
\begin{document}
\title{Synchronization in driven versus autonomous
coupled chaotic maps}
\author{M. Pineda$^1$ and   M. G. Cosenza$^2$}
\affiliation{$^1$Max-Planck-Institut f\"ur Physik Komplexer Systeme,
            N\"othnizer Strasse 38, 01187 Dresden, Germany\\
$^2$Centro de F\'{\i}sica Fundamental,
Universidad de Los
Andes,  Apartado Postal 26, M\'erida~5251, Venezuela.}
\begin{abstract}
The phenomenon of synchronization  occurring in a locally coupled map lattice subject to an
external drive is compared to the synchronization process in an autonomous coupled map system with similar local
couplings plus a global interaction. It is shown that chaotic synchronized states in both systems are equivalent, but the collective states arising after the chaotic synchronized state becomes unstable can be different in these two systems. It is found that the external drive induces chaotic synchronization  as well as synchronization of unstable periodic orbits of the local dynamics in the driven lattice.  On the other hand, the addition of a global interaction in the autonomous system allows for chaotic synchronization that is not possible in a large coupled map system possessing only local couplings.
 \end{abstract}
\maketitle

Coupled map lattices have provided fruitful theoretical and
computationally efficient models for investigating  a variety of
processes in spatially distributed dynamical systems, such as
synchronization, pattern formation, phase separation, turbulence,
nontrivial collective behavior,  etc. \cite{Chaos} . Recently, the
phenomena of synchronization  and pattern formation  induced by
external forcing on spatiotemporal dynamical systems, such as
chemical reactions \cite{Swinney,Hudson1,Kapral} have been the
focus of much attention. There has also been interest in
experimental investigations of spontaneous pattern formation and
emergence of synchronization in spatially extended systems of
interacting dynamical elements, such as one-dimensional arrays of
electrochemical oscillators \cite{Ring}, chemical and
hydrodynamical systems with global coupling \cite{Hudson2,Yamada},
and populations of chaotic electrochemical cells having both local
and global interactions \cite{Hudson3}. The relationship between
forced spatiotemporal systems and autonomous dynamical systems
possessing global interactions has recently been explored in the
framework of coupled map lattices \cite{PC}. In this paper, we
investigate the emergence of synchronization in forced
spatiotemporal systems by using a model of a coupled chaotic map
lattice subjected to an external drive. We show that the chaotic
synchronized state in this lattice is analogous to the chaotic
synchronized state emerging in an autonomous coupled map system
having similar  local couplings plus a global interaction, but
other collective states occur differently in these two systems. We
also show that the addition of a global interaction in the
autonomous system allows for chaotic synchronization that is not
possible in a large coupled map system possessing only local
couplings.

As a model of a driven spatiotemporal system, we consider a
one-dimensional coupled map lattice subjected to
a uniform external drive \cite{PC},
\begin{equation}
\label{drnw}\begin{array}{rl}
x^i_{t+1} & =(1-\epsilon_2)f(x_t^i)    \\
              & \\
                &  + \frac{\epsilon_1}{2}\left[(f(x_t^{i+1})+f(x_t^{i-1})-2f(x_t^i)\right] 
              +\epsilon_2 F_t \,.
\end{array}
\end{equation}
 where $x_t^i$ is the state of element $i$,
$(i=1,2,\ldots,N)$ at discrete time $t$; $N$ is the number of oscillators;
$f(x)$ describes the  local chaotic  dynamics;
 $\epsilon_1$ measures the local diffusive coupling,
$\epsilon_2$ expresses the coupling to the external
forcing, and $F_t$ is the uniform driving term which can be
any function of time.

The  dynamics of the  driven lattice can be compared with that of
an autonomous spatiotemporal system, described by
the following coupled map system possessing both, local and
global interactions,
\begin{equation}
\label{stcm}\begin{array}{rl}
x_{t+1}^i  & = (1-\epsilon_2)f(x_t^i)  
               +\frac{\epsilon_1}{2} \left[(f(x_t^{i+1})+f(x_t^{i-1}) \right.\\
              & \\
            &   \left. -2f(x_t^i) \right]
        +  \frac{\epsilon_2}{N}\sum_{i=1}^N f(x_t^i), 
\end{array}
\end{equation}
 where $f(x)$ is the same  local map as in  Eq.~(\ref{drnw});
$\epsilon_1$  and $\epsilon_2$ are the local and global coupling
parameters, respectively; and the global interaction is provided by  the mean field of the system.
Periodic boundary conditions are assumed in  both systems, Eq. (\ref{drnw}) and Eq.~(\ref{stcm}) .

When the
autonomous coupled map system in Eq.~(\ref{stcm}) reaches a
synchronized state at some parameter values, the evolution of its mean field
is identical to the dynamics of any local map. Thus for the same set of parameters,
the  driven
lattice, Eq.~(\ref{drnw}), subjected to a forcing $F_t$ equal to the local map should
exhibit a synchronized state similar to that of the associated autonomous coupled
map system Eq.~(\ref{stcm}) .

The driven lattice, Eq.~(\ref{drnw}),  can be expressed in vector form as
\begin{equation}
\label{vector}
{\bf x}_{t+1}=\left[ (1-\epsilon_2) I +  \frac{\epsilon_1}{2} L \right] {\bf f}({\bf  x}_{t})+ \epsilon_2{\bf  F}_{t},
\end{equation}
where  the vector components are  $\left[ {\bf  x}_{t}\right] _{i}=x^{i}_{t} $,
$  \left[ {\bf f}({\bf  x}_{t})\right] _{i}=f(x^{i}_{t}) $  and $ [{\bf F}_t]_i=F_t$;  $I$  is the $N \times N$
identity matrix,
and $L$ is the $N \times N$ matrix expressing the diffusive coupling among the elements, with components given
by $L_{i \, i\pm1}=1$, $L_{ii}=-2$,
and $L_{i j}=0$, otherwise.

The driven lattice may reach different asymptotic spatiotemporal patterns, ordered, synchronized, or turbulent,
depending on the
characteristics of the drive function $F_t$,  and on the initial conditions \cite{PC}. Here we consider
synchronized states of the lattice  induced by a periodic or a chaotic drive.
A synchronized state  at time $t$ is defined by  the condition $ x^i_t=x_t, \, \forall i$.
In a synchronized state,
the driven system
must satisfy
\begin{equation}\label{period}
x_{t+1}= \left( 1-\epsilon_2\right) f(x_t)  +\epsilon_2 F_t .
\end{equation}

The linear stability analysis of the synchronized state in the driven system leads to
the condition
\begin{equation}\label{stabdriven}
\left|\left( 1-\epsilon_2 +  \frac{\epsilon_1}{2} \nu_j\right) e^{\lambda}\right|< 1 \, ,
\end{equation}
where
$\nu_j= -4\sin ^{2}(\pi j/N) $, $j=0,1,2,\ldots, N-1$,  is the set of eigenvalues of the coupling matrix $L$,
with $N/2$ distinct eigenvalues and each being doubly degenerate \cite{Waller};
and $\lambda=\lim_{T\rightarrow\infty}\frac{1}{T}\sum_{t=0}^{T-1}\log \vert f'(x_t)\vert$ is the Lyapunov exponent
of the local map.
Thus the range of the  parameter $\epsilon_2$
for which a synchronized state is stable corresponds to
\begin{equation}\label{stable-unstable}
    1- 2\epsilon_1 \sin ^{2}(\frac{\pi  j}{N})-e^{-\lambda} < \epsilon_2
    < 1 -2\epsilon_1 \sin ^{2}(\frac{\pi j}{N})+ e^{-\lambda}.
\end{equation}

In particular, in a synchronized periodic  state all the elements  follow the same cyclic sequence of values.
Consider, for example, an  orbit of period $p$ of the local map, defined by
$f^{(p)}(\bar{x}_n)=\bar{x}_n$,
where $ \lbrace \bar{x}_1,\bar{x}_2 ,\ldots,\bar{x}_p \rbrace$ are the set of consecutive points belonging to the
orbit, satisfying $f(\bar{x}_n)=\bar{x}_{n+1}$, $ f(\bar{x}_p)=\bar{x}_1$. This periodic orbit is unstable if
$ e^{\lambda}=\prod^{p}_{n=1} \vert f'(\bar{x}_n)\vert > 1$.
If an unstable periodic orbit  gets synchronized in the driven lattice, then $x_{t+1}=\bar{x}_{n+1}$ and
$x_t=\bar{x}_n$, and
Eqs.~(\ref{period})
yield the solution $F_t=\bar{x}_{n+1}$. Thus, if the external drive follows a periodic
unstable orbit of the local map,  i.e., $F_t= \{\bar{x}_1,\bar{x}_2,\ldots,\bar{x}_p\}$,  then it
is possible to synchronize the entire  lattice on that orbit.

As an application, we shall consider a logarithmic  map  \cite{Kawabe}
$f(x)=b-\ln \vert  x  \vert$, $x \in (-\infty,\infty)$,  as local chaotic
dynamics in both Eq. (\ref{drnw}) and  Eq.~(\ref{stcm}).
This map exhibits robust chaos, with no periodic windows and no separated chaotic bands, in the
parameter interval $b \in [ -1,1]$.

In order to characterize the collective states in both systems, we calculate the mean field
\begin{equation}\label{mean}
h_t=\frac{1}{N}\sum_{i=1}^{N} f(x_t^i).
\end{equation}
Figures 1(a)-1(c) show  bifurcation diagrams of $h_t$ as a function of the coupling drive parameter
 $\epsilon_2$ for  lattices driven with different forms of $F_t$. For each value of $\epsilon_2$,
$h_t$  was calculated at each time step during a run starting from random initial conditions on the local maps,
uniformly distributed on the interval $[-8,4]$, after discarding the transients.
Figure 1(a) shows the bifurcation diagram of $h_t$ vs.
 $\epsilon_2$ for a lattice driven with a constant term $F_t=\bar{x}_1=-0.855$, where $\bar{x}_1=f(\bar{x}_1)$
is the unstable fixed point of the local logarithmic map for $b=-0.7$.
In the region labeled  by PS (periodic synchronization),
$h_t$ becomes equal to $\bar{x}_1$, indicating that the elements in the lattice are synchronized at this stationary value. The range of $\epsilon_2$ for which synchronization is observed corresponds to the range predicted by the stability condition Eq.~(\ref{stable-unstable}), for the unstable fixed point of $f$ with $p=1$ .
Outside the region of stationary synchronization, other types of collective behaviors can be observed in
the bifurcation diagram of Fig. 1(a). In the left region labeled by CPB (collective periodic behavior), the mean field of the  system driven with constant $F_t=\bar{x}_1$ experiences a collective period-$2$ motion, although the local elements are chaotic and desynchronized.
A collective fixed point occurs on the right CPB  region.
There is also a region, labeled by T (turbulence), where $h_t$ follows a normal statistical behavior around a
mean value with fluctuations reflecting the averaging of
$N$ completely desynchronized chaotic elements \cite{Gallego}.
\begin{figure}
\includegraphics[width=0.6\linewidth,angle=90]{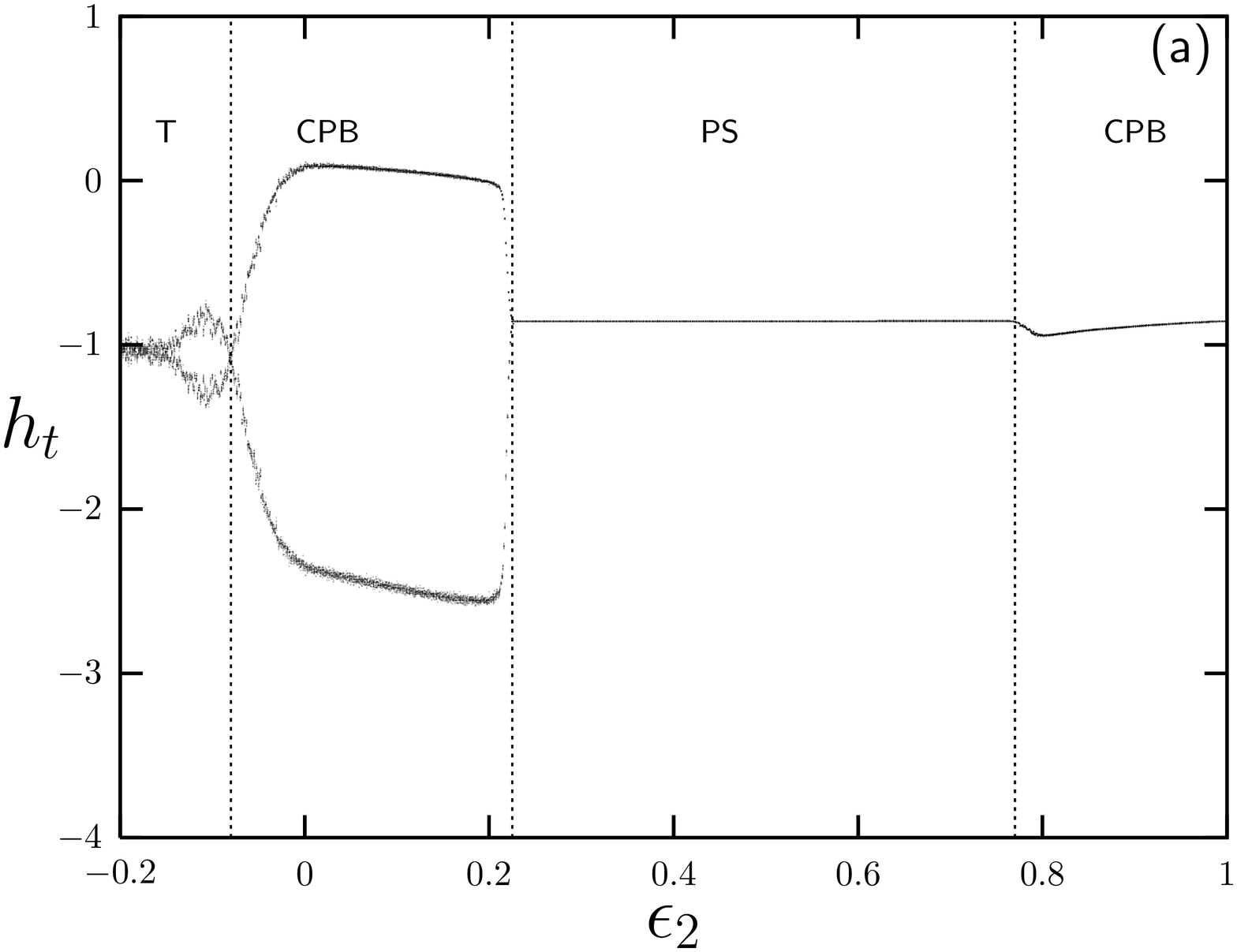}
\includegraphics[width=0.6\linewidth,angle=90]{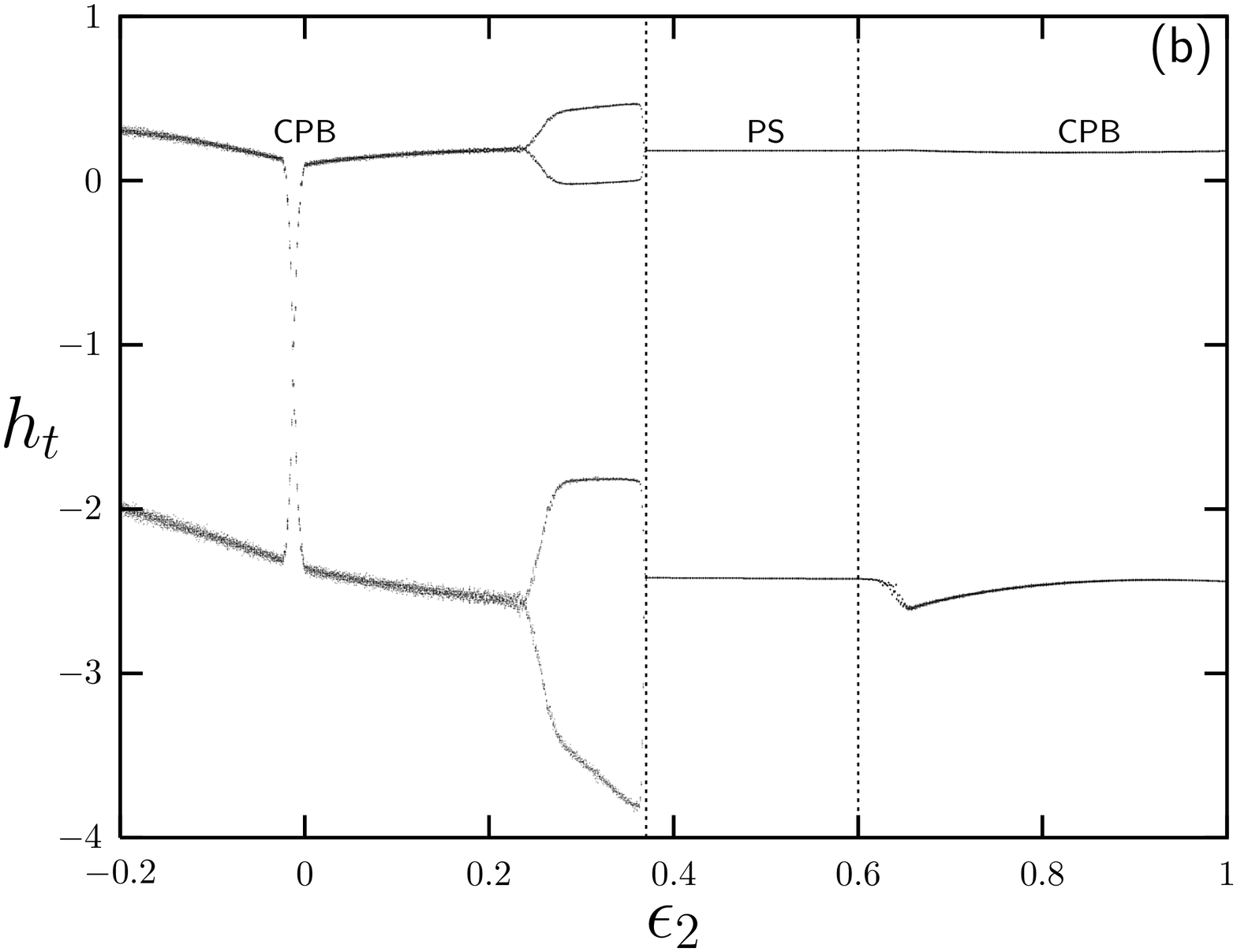}
\includegraphics[width=0.6\linewidth,angle=90]{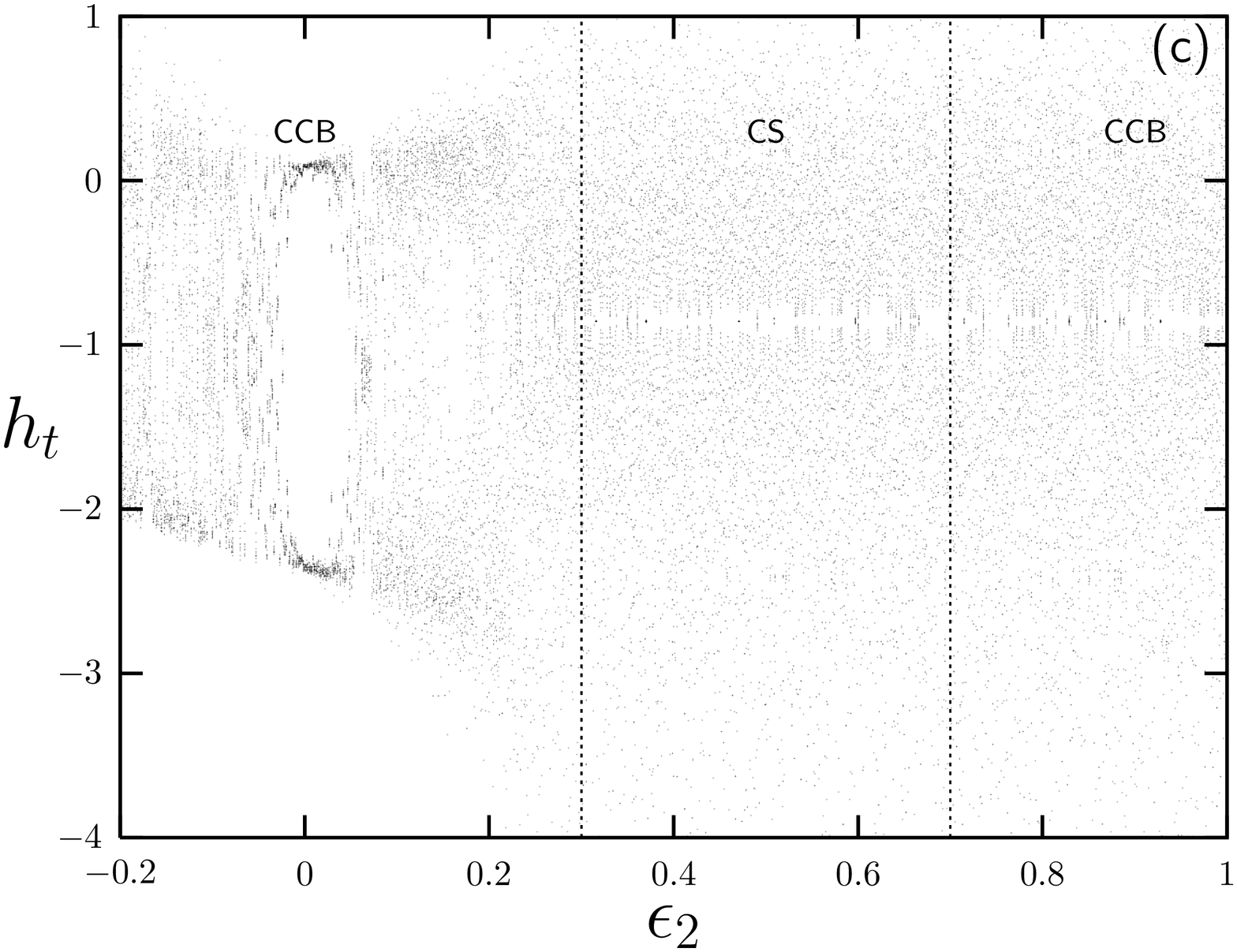}
\caption{Bifurcation diagrams of $h_t$ vs.
$\epsilon_2$ for the
driven lattice,  Eq.~(\ref{drnw}), with
$N=10^4$,
$\epsilon_1=0.5$,
$b=-0.7$. Labels indicate PS (periodic synchronization), CPB (collective periodic behavior), CS (chaotic synchronization), CCB (collective chaotic bands),
and T (turbulence).
(a) $F_t=\{\bar{x}_1=-0.855\}$.
(b)  $F_t=\{\bar{x}_1=0.18, \bar{x}_2=-2.44\}$.
(c)  $F_t=-0.7+\ln \vert x_t \vert$. }
\end{figure}
Figure 1(b) shows $h_t$ vs. $\epsilon_2$ for the lattice driven periodically with $F_t=\{\bar{x}_1,\bar{x}_2\}$, where
$\bar{x}_1=0.18$, $\bar{x}_2=-2.44$ are the points of the unstable period-$2$ orbit of the local logarithmic map for $b=-0.7$. In the region labeled by PS, the mean field coincides with this unstable periodic orbit, indicating that the lattice is synchronized on this orbit.  This range of $\epsilon_2$  is predicted by condition  Eq.~(\ref{stable-unstable}) for the unstable periodic orbit of $f$ with $p=2$ for the parameter value $b=-0.7$. Periodic collective  behaviors with period $2$ and period $4 $ in $h_t$ arise in the regions labeled by CPB.   Figure~1(c) corresponds to a chaotic driving, with $F_t=-0.7+\ln \vert x_t \vert$. Chaotic synchronization of the system occurs in the region labeled by CS and it is also predicted by Eq.~(\ref{stable-unstable} ).  In this CS region,  $h_t=F_t=-0.7+\ln \vert x_t \vert=x_t^i, \, \forall i$.   After crossing the boundaries of the CS region, the  collective states described by $h_t$ take the form of
chaotic bands. These states are labeled CCB (collective chaotic bands) and they consist of the motion of chaotic
elements that maintain some coherence.

Synchronized
states, $ x^i_t=x_t, \, \forall i$, can also emerge in the autonomous coupled map system Eq.~(\ref{stcm}). In order
to study the stability of these states, we  express
Eq.~(\ref{stcm}) in vector form as
\begin{equation}
\label{vector2}
{\bf x}_{t+1}=\left[ (1-\epsilon_2) I +  \frac{\epsilon_1}{2}L+\frac{\epsilon_2}{N}G \right]{\bf f}({\bf  x}_{t}),
\end{equation}
where the local connectivity matrix $L$ was defined in Eq.~(\ref{vector}) and $G$ is the $N \times N$ global
connectivity matrix  having all its components equal to $1$.

The linear stability analysis of synchronized states in the autonomous system  yields
\begin{equation}\label{range-driven}
\left|(1-\epsilon_2+ \frac{\epsilon_1}{2} \nu_j+\frac{\epsilon_2}{N} \gamma_j)e^{\lambda}\right|<1,
\quad
j=0,1,...,N-1,
\end{equation}
where $\nu_j$ are the eigenvalues of the local coupling matrix $L$ defined above,  and $\gamma_j=\delta_{oj}N$
correspond to the set of eigenvalues of $G$,  with the zero eigenvalue having $(N-1)$-fold degeneracy.
The range of $\epsilon_2$ for which synchronization takes place is
\begin{equation}\label{auto}
    1- 2\epsilon_1 \sin ^{2}(\frac{\pi j}{N})-e^{-\lambda} < \epsilon_2
    < 1 -2\epsilon_1 \sin ^{2}(\frac{\pi j}{N})+ e^{-\lambda}  ,
\end{equation}
which is the same condition for stability of synchronized states
in the driven lattice, Eq.~(\ref{stable-unstable}). However, the
unstable periodic orbits of the local map $f$ can not be
synchronized in the autonomous system because they correspond to
unstable synchronized states in this system.

The eigenvector corresponding to $j=0$ is  homogeneous for both
matrices $L$ and $G$. Thus perturbations of ${\bf x}_t$ along this eigenvector do not destroy the coherence,
and the stability condition associated with $j=0$ is irrelevant for a synchronized state. The other eigenvalues
corresponding to $j \neq 0$ are not homogeneous. Thus, condition Eq.~(\ref{stable-unstable} ) with $j \neq 0$
define  regions in the space of parameters where all the stable synchronized states can be observed.

Figure~2 shows the boundary curves given by
Eq.~(\ref{range-driven}) on the parameter  plane
$(\epsilon_2,\epsilon_1)$, in the limit of large $N$. The label CS
indicates where a chaotic synchronized state is stable in the
autonomous system. After crossing the stability boundaries of the
CS state, the autonomous system exhibits spatiotemporal patterns
corresponding to short wave (SW) and long wave (LW) modes, as
indicated in Fig.~2. Note  that the stability condition
Eq.~(\ref{range-driven}) does not hold  for $\epsilon_2=0$ and
large $N$; i.e.,  chaotic synchronization for large system size
$N$ can not take place in coupled map lattices possessing  only
local couplings, which is a well known fact \cite{Pikovsky}. Thus,
the addition of a global interaction allows the emergence of
chaotic synchronization in a large locally coupled system.
Similarly, an external uniform driving can be used to induce
chaotic synchronization in a locally coupled map network.
\begin{figure}[htb]
\includegraphics[width=0.6\linewidth,angle=90]{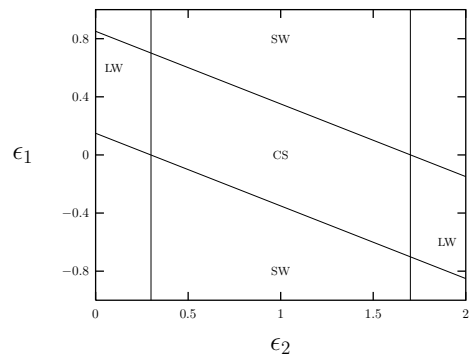}
\caption{Stability boundaries of the chaotic synchronized state CS of the autonomous
system Eq.~(\ref{stcm})  on the parameter  plane $(\epsilon_2,\epsilon_1)$,
with  $N=10^4$, $b=-0.7$.  Regions corresponding to short wave (SW) and long wave (LW)  patterns emerging
from the CS state are indicated.}
\end{figure}

Figure 3(a) shows the bifurcation diagram of $h_t$ for the autonomous system, Eq.~(\ref{stcm}), as a function of
the coupling parameter $\epsilon_2$.  In contrast to the behavior displayed by the driven lattice,  no regions of
synchronization of unstable periodic orbits of the local dynamics in the space of parameters are observed in the autonomous system, as expected. The range of chaotic synchronization (CS) corresponds to the same range of $\epsilon_2$ for chaotic synchronization in Fig.~1(c) for the driven lattice.  Note, however, that beyond the region CS in Fig.~3(a),  there are collective states emerging in the autonomous system, such as CPB (collective periodic behavior),
and T (turbulence),  that do not appear in the corresponding diagram of the driven lattice, Fig.~1(c).
By varying the coupling strengths, various spatial patterns can be realized in the
autonomous system. These patterns correspond
to short and to long wave modes and can be regarded as generalized Turing patterns that emerge
when a global synchronized state becomes unstable \cite{Ding}.
Figure~3(b) shows the bifurcation diagram of $h_t$ vs. $\epsilon_1$, with  $\epsilon_2$ fixed.
\begin{figure}
\includegraphics[width=0.6\linewidth,angle=90]{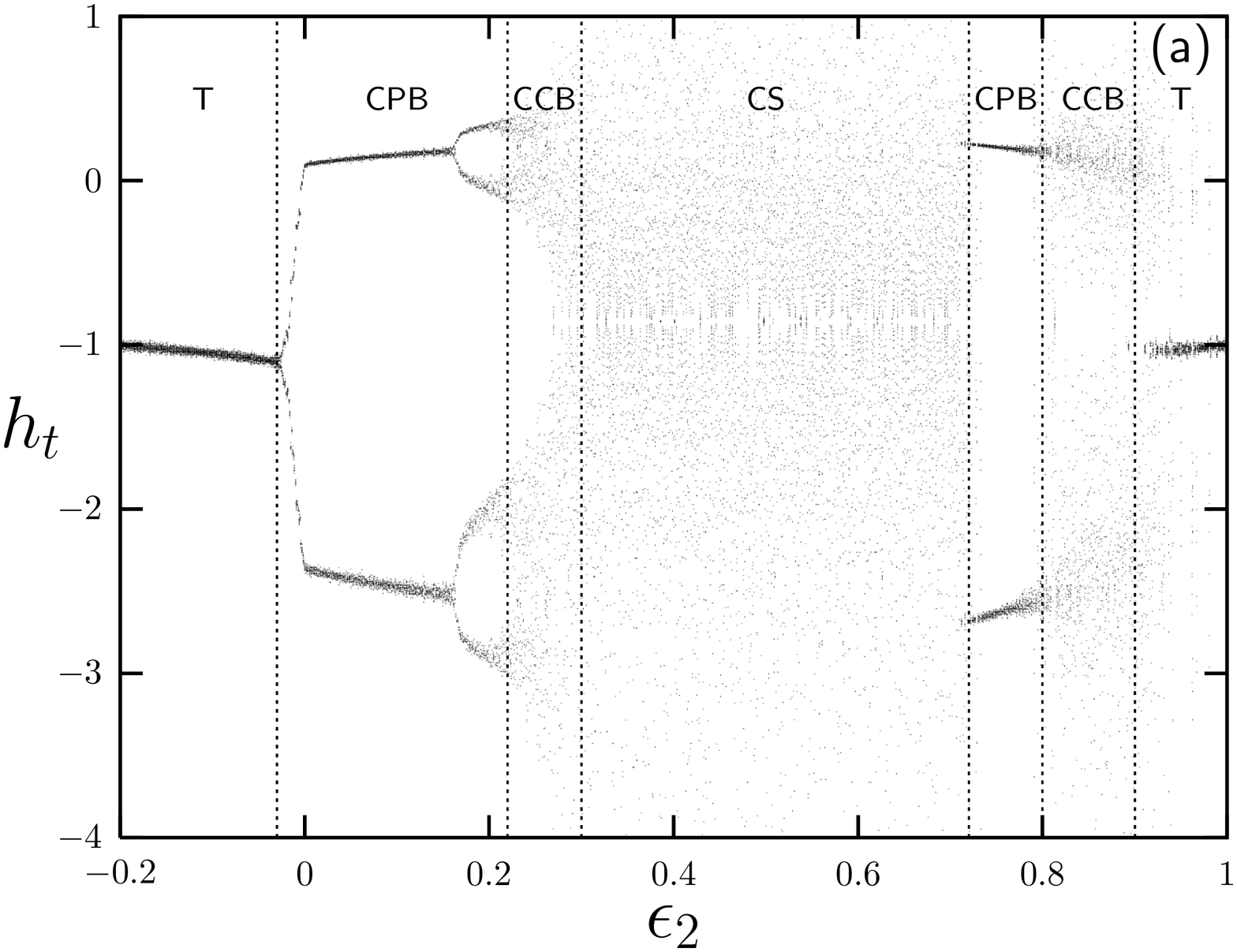}
\includegraphics[width=0.6\linewidth,angle=90]{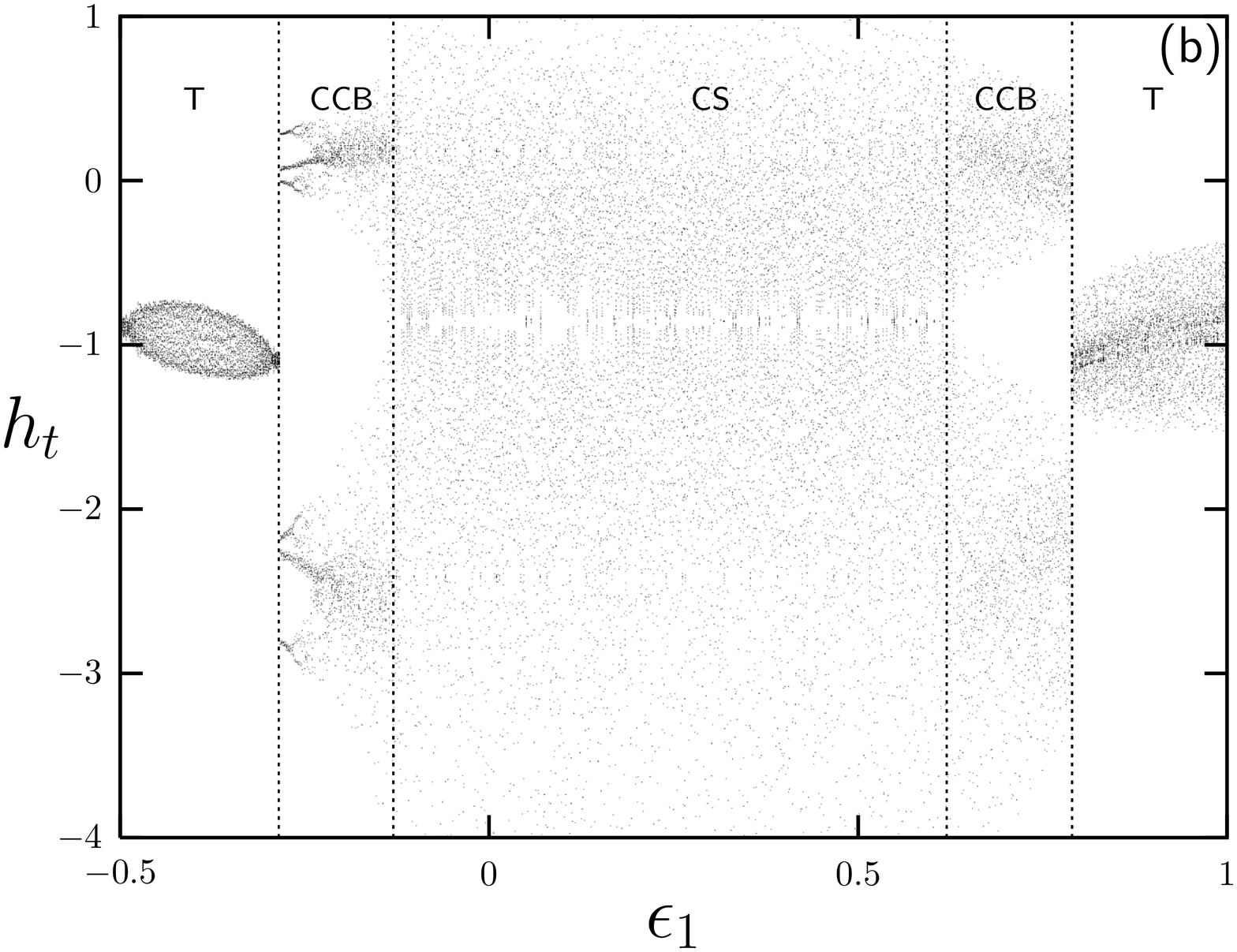}
\caption{Bifurcation diagrams of
$h_t$ for the autonomous system,  for $b=-0.7$, $N=10^4$. (a) $h_t$ vs. $\epsilon_2$,
with fixed $\epsilon_1=0.5$. (b) $h_t$ vs. $\epsilon_1$,  with fixed $\epsilon_2=0.5$.   Labels are defined in Fig.~1}
\end{figure}

The collective states observed in the bifurcation diagrams of Fig.~3, as well as those in Fig.~1, have been checked for system-size effects.  When the lattice size $N$ is increased, the segments  in the CPB regions shrink,
but the amplitudes of the collective periodic motions
do not decrease,
indicating that the global periodic attractors become better defined in the large system limit.
As a consequence, the variance of the fluctuations of $h_t$ itself do not decrease
as $N^{-1}$ with increasing $N$,
but rather it saturates at some constant value related to the amplitude of the collective periodic motion.
These states of  ordered  evolution in macroscopic quantities such as $h_t$, arising from local chaos in spatiotemporal
systems,  have been called nontrivial collective behavior \cite{Chate}.
The states of collective chaotic bands (CCB) are also manifestations of nontrivial collective behavior, since the variance of $h_t$ for those states neither follows a regular statistical behavior.
In contrast, the variance of the mean field in the turbulent states (T) appears to decrease as $N^{-1}$ with
increasing $N$, obeying normal statistical behavior.

In summary, we have studied the analogy  between the chaotic
synchronized states emerging in forced spatiotemporal systems and
in autonomous dynamical systems possessing global interactions in
the context of coupled map lattices. By using a model of a
one-dimensional coupled map lattice subjected to a uniform
external drive, we have shown that both, synchronization of
unstable periodic orbits of the local maps and chaotic
synchronized states,  can be induced in the driven lattice. The
external drive acts as a control mechanism for stabilizing
unstable periodic orbits of the local maps. We showed that the
synchronization behavior of the driven lattice can be compared,
under some circumstances, with that of an autonomous coupled map
system possessing a similar local coupling plus an additional
global interaction that acts as a global feedback. It was found
that the chaotic synchronized states in both systems are
analogous; however, the autonomous system does not exhibit
synchronization of unstable local periodic orbits. The collective
states arising when the chaotic synchronized state becomes
unstable are, in general,  different in these systems. Although we
have considered one-dimensional diffusive couplings, expressed by
the matrix $L$, the analogy between a uniform external drive and a
global interaction can also be applied to other networks of
coupled maps whose connectivity may be represented in matrix form.

M.P. is supported by a DAAD scholarship, Germany.
We thank H. Kantz and A.E. Motter for useful comments.


\begin{thebibliography}{99}
\bibitem{Chaos} Chaos {\bf 2}, No. 3 (1992), focus issue on
Coupled Map Lattices; edited by K.Kaneko.
\bibitem{Swinney} A. L. Lin, A. Hagberg, A. Ardelea, M. Bertram,
H. L. Swinney, and E. Meron, Phys. Rev. E {\bf 62}, 3790 (2000).
\bibitem{Hudson1} W. Wang, I. Z. Kiss, and J. L. Hudson,
Phys. Rev. Lett. {\bf 86}, 4954 (2001).
\bibitem{Kapral} C. Hemming and R. Kapral, Physica A {\bf 306}, 199 (2002).
\bibitem{Ring} W. Wang, I. Z. Kiss, and J. L. Hudson,
J. Phys. Chem. B {\bf 103} 2178 (1999).
\bibitem{Hudson2} W. Wang, I. Z. Kiss, and J. L. Hudson,
Chaos {\bf 10} 248 (2000).
\bibitem{Yamada} K. Miyakawa and K. Yamada, Physica D, {\bf 151}, 217 (2001).
\bibitem{Hudson3} I. Z. Kiss, Y. Zhai, and J. L. Hudson,
Phys. Rev. Lett. {\bf 88}, 238301 (2002).
\bibitem{PC}  M. G. Cosenza, M. Pineda, and A. Parravano, Phys. Rev. E {\bf 67},
066217 (2003).
\bibitem{Waller} I. Waller and R. Kapral, Phys. Rev. A {\bf 30}, 2047 (1984).
\bibitem{Kawabe} T. Kawabe and K. Kondo, Prog. Theor. Phys.  {\bf 85}, 759 (1991).
\bibitem{Gallego} M. G. Cosenza and J. Gonz\'alez, Prog. Theor. Phys. {\bf 100}, 21 (1998).
\bibitem{Pikovsky} A. Pikovsky, M. Rosenblum, and J. Kurths, Synchronization, a universal concept
in nonlinear sciences, Cambrigde University Press, Cambridge, UK   (2001).
\bibitem{Ding} G. Rangarajan, Y. Chen, and M. Ding, Phys. Lett. A {\bf 310}, 415 (2003).
\bibitem{Chate} H. Chat\'e and P. Manneville, Prog. Theor. Phys. {\bf  87}, 1 (1992).
\end{thebibliography}
\end{document}